\documentclass[12pt]{article}

\usepackage{scicite}
\usepackage{amsmath}
\usepackage{hyperref}
\usepackage{graphicx}

\usepackage{times}

\newenvironment{sciabstract}{%
\begin{quote} \bf}
{\end{quote}}

\newcounter{lastnote}

\title{Spiral Density Waves in a Young Protoplanetary Disk}

\author
{Laura M. P\'erez,$^{1\ast}$ John M. Carpenter,$^{2}$ Sean M. Andrews,$^{3}$ Luca Ricci,$^{3}$ 
\\
Andrea Isella,$^4$ Hendrik Linz,$^5$ Anneila I. Sargent,$^6$ David J. Wilner,$^3$ \\
Thomas Henning,$^5$ Adam T. Deller,$^7$ Claire J. Chandler,$^8$ \\
Cornelis P. Dullemond,$^9$ Joseph Lazio,$^{10}$ Karl M. Menten,$^1$ \\
Stuartt A. Corder,$^2$ Shaye Storm,$^3$ Leonardo Testi,$^{11,12}$ Marco Tazzari,$^{11}$ \\
Woojin Kwon,$^{13,14}$
Nuria Calvet,$^{15}$ Jane S. Greaves,$^{16}$\\
Robert J. Harris,$^{17}$ Lee G. Mundy$^{18}$
\\
\\
\normalsize{$^{1}$Max-Planck-Institut f\"ur Radioastronomie, Bonn, Germany,}\\
\normalsize{$^{2}$Joint ALMA Observatory,  Vitacura, Santiago, Chile,}\\
\normalsize{$^{3}$Harvard-Smithsonian Center for Astrophysics, Cambridge, MA, USA,}\\
\normalsize{$^{4}$Rice University,  Houston, TX, USA,}\\
\normalsize{$^{5}$Max-Planck-Institut f\"ur Astronomie, Heidelberg, Germany,}\\
\normalsize{$^{6}$California Institute of Technology, Pasadena, CA, USA,}\\
\normalsize{$^{7}$The Netherlands Institute for Radio Astronomy (ASTRON), The Netherlands,}\\
\normalsize{$^{8}$National Radio Astronomy Observatory, Socorro, NM, USA,}\\
\normalsize{$^{9}$Heidelberg University, Center for Astronomy, Heidelberg, Germany,}\\
\normalsize{$^{10}$Jet Propulsion Laboratory, California Institute of Technology, Pasadena, CA, USA}\\
\normalsize{$^{11}$European Southern Observatory, Garching, Germany,}\\
\normalsize{$^{12}$INAF-Osservatorio Astrofisico di Arcetri, Firenze, Italy,}\\
\normalsize{$^{13}$Korea Astronomy and Space Science Institute, Daejeon, Republic of Korea,}\\
\normalsize{$^{14}$Korea University of Science and Technology, Daejeon, Republic of Korea,}\\
\normalsize{$^{15}$University of Michigan, Ann Arbor, MI, USA,}\\
\normalsize{$^{16}$Cardiff University, School of Physics \& Astronomy,Cardiff, UK,}\\
\normalsize{$^{17}$University of Illinois, Urbana, IL, USA,}\\
\normalsize{$^{18}$Department of Astronomy, University of Maryland, College Park, MD, USA.}\\
\normalsize{$^\ast$To whom correspondence should be addressed; E-mail: lperez@mpifr-bonn.mpg.de}
}

\date{}

\begin{document} 


\baselineskip16pt


\maketitle


\begin{sciabstract}
  Gravitational forces are expected to excite spiral density waves in protoplanetary disks, disks of gas and dust orbiting young stars. However, previous observations that showed spiral structure were not able to probe disk midplanes, where most of the mass is concentrated and where planet formation takes place. Using the Atacama Large Millimeter/submillimeter Array we detected a pair of trailing symmetric spiral arms in the protoplanetary disk surrounding the young star Elias 2-27. The arms extend to the disk outer regions and can be traced down to the midplane. These millimeter-wave observations also reveal an emission gap closer to the star than the spiral arms. We argue that the observed spirals trace shocks of spiral density waves in the midplane of this young disk.
\end{sciabstract}



Spiral density waves are expected to be excited in the midplane of protoplanetary disks by the action of gravitational forces, generated for example by planet-disk interactions ({\it 1}) or by gravitational instabilities ({\it 2}). These waves give rise to spiral structure whose observable characteristics – the number and location of arms, their amplitudes and pitch angles – depend on the driving mechanism and the disk physical properties ({\it 1,3–-5}). Theoretical predictions agree that these spiral features can be very prominent and thus more easily observable than the putative embedded planets or instabilities driving such waves ({\it 6,7}). Spiral-like patterns have been observed in evolved protoplanetary disks with depleted inner regions, in optical scattered light ({\it 8–-13}) or gas spectral lines ({\it 14,15}). However, at the wavelength of such observations the emission is optically thick and scattered light only traces the tenuous surface layers of these disks rather than their midplane densities. This makes it impossible to disentangle between minute perturbations near the disk surface and true density enhancements over the disk column density due to spiral density waves ({\it 16,5}). To probe the disk density structure, particularly the disk midplane that contains most of the mass and where planets form, observations of optically thin emission are necessary. 

We used the Atacama Large Millimeter/submillimeter Array (ALMA) to observe the protoplanetary disk around the young star Elias 2-27 at a wavelength of 1.3 mm. Our spatially resolved image (Fig.\ 1) shows two symmetric spiral arms extending from an elliptical emission ring. To emphasize the spirals and the dark ring of attenuated emission seen at $\approx70$ AU radius, we applied an unsharp masking filter ({\it 17}) to increase significantly the image contrast (Fig.\ 1B).

The young star Elias 2-27 ({\it 18}) is a member of the $\rho$-Ophiuchus star-forming complex at a distance of 139 pc ({\it 19}) and is classified as a Class II young stellar object from analysis of its spectral energy distribution (SED, {\it 20,21}). Although the star is only 50-60\% of the Sun's mass ($M_{\odot}$) ({\it 20,22}) it is known to harbor an unusually massive (0.04-0.14 $M_{\odot}$, {\it 20,23,24}) protoplanetary disk. The star, obscured by 15 magnitudes of extinction at optical wavelengths by the parent molecular cloud ({\it 22}), accretes material from its surrounding disk with a mass accretion rate of $8\times10^{-8}$ $M_{\odot}\; yr^{-1}$ ({\it 25}). Previous observations at $0.6''$-$1.1''$ angular resolution were well described by a smooth and axisymmetric distribution of material in the disk that extends near the stellar photosphere and decreases monotonically with distance from the star ({\it 20,23}).

To estimate the optical depth of the observed dust continuum emission, we performed radiative transfer calculations using RADMC3D ({\it 26}) at 1.3mm ({\it 17}) using the previous surface density constraint found for the Elias 2-27 disk ({\it 20}). This model reproduces the azimuthally averaged radial profile of the observed ALMA 1.3 mm continuum emission ({\it 17}, Fig.\ S1). At a radial distance from the star (hereafter referred to as radius and denoted by $R$) larger than $R\approx10$~AU the emission is optically thin and thus traces the density of solid material down to the midplane of the disk. At the location of the spiral structures (from $R=100$ to $300$~AU) the azimuthally averaged optical depth $\tau$ of the dust continuum emission is $\tau=0.1$ at $R=100$~AU, decreasing to $\tau=0.02$ at $R=300$~AU (Fig.\ S1B), consistent with the measured peak brightness temperature on the spirals of 1.2 K at $R=150$~AU.

The spiral structures are even more evident in Figure 2A, where the data has been projected into a polar coordinate grid that accounts for the viewing geometry of the disk. In polar coordinates, a ring with zero eccentricity would have a constant radius for all polar angles. However, Figure 2A shows two bright structures that grow in radius from about 100~AU to 300~AU as the polar angle increases. The brightest of these two structures lies Northwest of the star, labeled NW, the spiral structure Southeast of the star is labeled SE. In Figure 2B we present the surface brightness contrast of the NW and SE arms, defined as the ratio between the peak of emission at the arm and the background surface brightness ({\it 17}). We find both arms have similar contrasts ranging between values of 1.3 and 2.5. The spiral arms reach their highest contrast at $R=150$~AU, coinciding with the location in the disk where gravity has the most influence over thermal pressure and shear forces, i.e. where the Toomre $Q$ parameter is lowest ({\it 17}, Fig.\ S2). However, even at its minimum value, Toomre $Q$ is well inside the stable regime ({\it 17}). If the spirals arms suffer from beam dilution (i.e. their physical size is smaller than the angular resolution of our observation) a higher optical depth than our previous estimate could be possible, implying an even higher density contrast in the arms. Thus, the contrast values measured for NW and SE are lower limits.

We determined the local maxima and minima of emission in the dust continuum observations at evenly spaced azimuthal angles, after subtracting a smooth monotonically decreasing intensity profile that best fit the intensity radial profile of the disk ({\it 17}, Fig.\ S3). Figure 3 demonstrates that the emission local maxima (crosses) describe two spiral structures, while the emission local minima (circles) describe an ellipse. We constrained the geometry of these structures by modeling their location in polar coordinates (where $R$ is the distance from the star located at the origin and $\theta$ the angle from the $x$-axis), taking into account that these structures have been inclined and rotated with respect to our line of sight by their inclination ($i$) and position angle ($PA$). The emission local minima were fitted with a circular ring ($R=a_0$, with $a_0$ the radius at which the gap is located), while the emission local maxima were fitted with two symmetric logarithmic spirals ($R=R_0 e^{b\theta}$, with $R_0$ the spiral radius at $\theta=0^{\circ}$ and $b$ the rate at which the spirals increase their distance from the origin). The best-fit parameters for the symmetric spirals that describe the local maxima are $R_0 = 84 \pm 4$~AU and $b = 0.138 \pm 0.007$ (i.e. a pitch angle of $\phi = 7.9^{\circ} \pm 0.4^{\circ}$), while the circular ring that describes the local minima has a radius of $a_0 = 71 \pm 2$~AU. The geometry of the spiral arms and dark ring can all be described with a single inclination angle of $i = 55.8^{\circ} \pm 0.9^{\circ}$ and position angle $PA = 117.3^{\circ} \pm 0.9^{\circ}$. Figure 3 shows the best-fit model and constraints at the $3\sigma$ level for the spiral arms and dark ring. 

Spatially resolved molecular line observations of CO and two isotopologues, simultaneous to the continuum observations discussed here, suggest that the South-West side of the disk is tilted toward Earth while the disk rotates in a clockwise direction in a Keplerian velocity pattern ({\it 17}, Fig.\ S4-S7). It is most likely that the observed NW and SE spirals point away from the direction of rotation, i.e. these are trailing spiral arms.

Simulations indicate that planet/companion-disk interactions (PDI) alone are capable of opening a gap creating density and temperature contrasts along spiral arms that are consistent with the observed values in NW and SE ({\it 1,6}). However, the observed gap at 70 AU is of low contrast, indicative of a narrow or partially opened gap as predicted from PDI with low-mass planets. Such planets will have a harder time exciting spiral density waves of the observed contrast in Elias 2-27 ({\it 27}). Grain growth could in principle create such a low-contrast gap at 70~AU, but we see no evidence of this process in our 9~mm observations from the Karl G. Jansky Very Large Array ({\it 17}, Fig.\ S8). Additionally, PDI simulations generally predict spiral arms with different contrasts to each other, unlike the similar contrast of NW and SE. Those models often struggle to produce the symmetric spiral pattern observed in the Elias 2-27 disk ({\it 27,28}), unless the planet-induced density waves are driven by a massive companion at large disk radii exterior to the spirals ({\it 6}), of which we see no evidence in the Elias 2-27 system. 

Alternatively, gravitational instabilities (GI) can also excite spiral density waves with contrasts in density and temperature that are similar to the observed contrasts in NW and SE. A high mass accretion rate, on the order of $10^{-6}-10^{-7}$ $M_{\odot}\; yr^{-1}$ ({\it 29}), coupled with a large disk-to-star mass ratio, of at least $M_{disk}$/$M_{star} \approx 0.5$ ({\it 3}), are required to induce the “grand design” symmetric spiral arms observed in Elias 2-27. Although the mass accretion rate of Elias 2-27 is high ({\it 25}), even in the most optimistic case the disk-to-star mass ratio is $M_{disk}$/$M_{star} \approx 0.3$, limiting the possibility of GI acting alone. Additionally, simulations of disks undergoing GI generally cannot maintain spiral arms at radii larger than 100~AU, as the disk begins to fragment at large radii ({\it 29--31,2,3}). A possibility to avoid fragmentation in the GI scenario is that the Elias 2-27 disk is in a marginally unstable state ({\it 32}), for example supported by external irradiation or sustained by an envelope that is feeding mass to the outer disk. However, no evidence for a massive envelope is found in the infrared SED of this object ({\it 20,21}), making the possibility of a marginally unstable disk unlikely. 

A combination of PDI and GI mechanisms acting together ({\it 5}) could be another alternative to explain the high degree of symmetry of the arms in both contrast and location, the presence of both arms out to large disk radii, together with the gap in the disk located at 70~AU. However, this scenario also requires a high-mass planet ({\it 5}), whose effect in the disk structure should be discernable as a larger and deeper gap than the one constrained in this work.

The observed structure in Figure 1 delineates two symmetric trailing spiral arms, NW and SE, whose low optical depth enables material to be traced in the disk midplane. Together with the measured contrast values of at least 1.3--2.5 in the spiral arms, these results imply that NW and SE are tracing density and/or temperature enhancements at the disk midplane, which are arranged into two “grand design” symmetric spiral arms. Given the disk differential rotation (similar to the case of spiral galaxies), if the observed spirals were material arms rotating at the disk Keplerian velocity then the inner part of the spiral (at 100~AU) would rotate 5 times faster than the outer part of the arm (at 300~AU). Thus, in a timescale much shorter than the lifetime of the disk (i.e.\ only a few orbital periods of roughly 1000~years) these spiral arms would wind-up and disappear. We conclude that instead of tracing material arms, the NW and SE spiral structures trace density and temperature enhancements due to spiral density waves in the midplane of the Elias 2-27 disk. 

Unlike the spiral features from scattered-light observations, the spiral arms detected with our millimeter-wave imaging trace structures at the midplane of the disk –-- where planet formation takes place –-- allowing us to discern the location, shape, contrast, and size of these spiral density waves at the disk midplane. These results provide a unique benchmark for numerical simulations of spiral structure in protoplanetary disks, particularly as fragmentation of such spirals remains the only plausible formation mechanism for planets/companions at large disk radii, where core-accretion becomes inefficient ({\it 33}). Thus, the detection of spiral features in Elias\,\,2-27 is a first step to determine what is the dominant mechanism of planet formation at different locations in the disk.

\subsection*{References and Notes}
\begin{enumerate}

\item Z. Zhu, R. Dong, J. M. Stone, R. R. Rafikov. The structure of spiral shocks excited by planetary-mass companions. Astrophys. J. 813, 88 (2015).
\item K. M. Kratter, G. Lodato. Gravitational instabilities in circumstellar disks. \url{http://arxiv.org/abs/1603.01280} (2016).
\item R. Dong, C. Hall, K. Rice, E. Chiang. Spiral arms in gravitationally unstable protoplanetary disks as imaged in scattered light. Astrophys. J. 812, L32 (2015).
\item M. Flock, J. P. Ruge, N. Dzyurkevich, T. Henning, H. Klahr, S. Wolf. Gaps, rings, and non-axisymmetric structures in protoplanetary disks. From simulations to ALMA observations. Astron. Astrophys. 574, A68 (2015).
\item A. Pohl et al. Scattered light images of spiral arms in marginally gravitationally unstable discs with an embedded planet. Mon. Not. R. Astron. Soc. 453, 1768 (2015).
\item R. Dong, Z. Zhu, R. R. Rafikov, J. M. Stone. Observational signatures of planets in protoplanetary disks: spiral arms observed in scattered light imaging can be induced by planets. Astrophys. J. 809, L5 (2015).
\item G. Dipierro, G. Lodato, L. Testi, I. de Gregorio Monsalvo. How to detect the signatures of self-gravitating circumstellar discs with the Atacama Large Millimeter/sub-millimeter Array. Mon. Not. R. Astron. Soc. 444, 1919 (2014).
\item M. Fukagawa et al. Near-infrared images of protoplanetary disk surrounding HD 142527. Astrophys. J. 636, L153 (2006).
\item J. Hashimoto et al. Direct imaging of fine structures in giant planet-forming regions of the protoplanetary disk around AB Aurigae. Astrophys. J. 729, L17 (2011).
\item S. P. Quanz, et al. Very Large Telescope/NACO polarimetric differential imaging of HD 100546 - disk structure and dust grain properties between 10 and 140 AU. Astrophys. J. 738, 23 (2011).
\item T. Muto, et al. Discovery of small-scale spiral structures in the disk of SAO 206462 (HD 135344B): implications for the physical state of the disk from spiral density wave theory. Astrophys. J. 748, L22 (2012).
\item C. A. Grady, et al. Spiral arms in the asymmetrically illuminated disk of MWC 758 and constraints on giant planets. Astrophys. J. 762, 48 (2013).
\item M. Benisty, et al. Asymmetric features in the protoplanetary disk MWC 758. Astron. Astrophys. 578, L6 (2015).
\item V. Christiaens, S. Casassus, S. Perez, G. van der Plas, F. M\'enard. Spiral arms in the disk of HD 142527 from CO emission lines with ALMA. Astrophys. J. 785, L12 (2014).
\item Y. W. Tang, S. Guilloteau, V. Pi\'etu, A. Dutrey, N. Ohashi, P. T. P. Ho. The circumstellar disk of AB Aurigae: evidence for envelope accretion at late stages of star formation? Astron. Astrophys. 547, A84 (2012).
\item A. Juh\'asz, M. Benisty, A. Pohl, C. P. Dullemond, C. Dominik, S.-J. Paardekooper. Spiral arms in scattered light images of protoplanetary discs: are they the signposts of planets? Mon. Not. R. Astron. Soc. 451, 1147 (2015).
\item Materials and methods are available as supplementary materials on Science Online and as an additional section below in this manuscript.
\item The star is also known as GSS 39, VSSG 28, GY116, and/or BKLT J162645.242309
\item E. E. Mamajek. On the distance to the Ophiuchus star-forming region. Astronomische Nachrichten 329, 10 (2008).
\item S. M. Andrews, D. J. Wilner, A. M. Hughes, C. Qi, C. P. Dullemond. Protoplanetary disk structures in Ophiuchus. Astrophys. J. 700, 1502 (2009).
\item N. J. Evans II, et al. The Spitzer c2d Legacy Results: Star-Formation Rates and Efficiencies; Evolution and Lifetimes. Astrophys. J. Suppl. Ser. 181, 2, 321-350 (2009).
\item A. Natta, L. Testi, S. Randich. Accretion in the ρ-Ophiuchi pre-main sequence stars. Astron. Astrophys. 452, 245 (2006).
\item A. Isella, J. M. Carpenter, A. I. Sargent. Structure and evolution of pre-main-sequence circumstellar disks. Astrophys. J. 701, 260 (2009).
\item L. Ricci, L. Testi, A. Natta, K. J. Brooks. Dust grain growth in ρ-Ophiuchi protoplanetary disks. Astron. Astrophys.  521, A66 (2010).
\item J. R. Najita, S. M. Andrews, J. Muzerolle. Demographics of transition discs in Ophiuchus and Taurus. Mon. Not. R. Astron. Soc. 450, 3559 (2015).
\item C. P. Dullemond. RADMC-3D: A multi-purpose radiative transfer tool. Astrophysics Source Code Library, record ascl:1202.015 (2012).
\item J. Fung, R. Dong. Inferring planet mass from spiral structures in protoplanetary disks. Astrophys. J. 815, L21 (2015).
\item J. Bae, Z. Zhu, L. Hartmann. Planetary signatures in the SAO 206462 (HD 135344B) disk: a spiral arm passing through vortex? Astrophys. J. 819, 134 (2016).
\item C. Hall, D. Forgan, K. Rice, T. J. Harries, P. D. Klaassen, B. Biller. Directly observing continuum emission from self-gravitating spiral waves. Mon. Not. R. Astron. Soc., 458, 306 (2016).
\item C. J. Clarke. Pseudo-viscous modelling of self-gravitating discs and the formation of low mass ratio binaries. Mon. Not. R. Astron. Soc. 396, 1066 (2009).
\item P. Cossins, G. Lodato, L. Testi. Resolved images of self-gravitating circumstellar discs with ALMA. Mon. Not. R. Astron. Soc. 407, 181 (2010).
\item G. Bertin, G. Lodato. A class of self-gravitating accretion disks. Astron. Astrophys. 350, 694-704 (1999).
\item K. Rice. The Evolution of Self-Gravitating Accretion Discs. Publications of the Astronomical Society of Australia, 33, e012 (2016).

\end{enumerate}

\noindent {\bf Acknowledgements} We thank L. Loinard for useful discussions. L.M.P. acknowledges support from the Alexander von Humboldt Foundation. A.I. acknowledges support from NSF award AST-1109334/1535809 and from the NASA Origins of Solar Systems program through award number NNX14AD26G. A.I.S. is partially supported by NSF grant AST 1140063. This paper makes use of the following ALMA data: ADS/JAO.ALMA\#2013.1.00498.S and it can be obtained from the ALMA Science Data Archive, \url{https://almascience.nrao.edu/alma-data/} (raw format) and in the calibrated fits format used for analysis here from \url{https://safe.nrao.edu/evla/disks/elias2-27/} . ALMA is a partnership of ESO (representing its member states), NSF (USA) and NINS (Japan), together with NRC (Canada), NSC and ASIAA (Taiwan), and KASI (Republic of Korea), in cooperation with the Republic of Chile. The Joint ALMA Observatory is operated by ESO, AUI/NRAO and NAOJ. The National Radio Astronomy Observatory is a facility of the National Science Foundation operated under cooperative agreement by Associated Universities, Inc. Part of this research was carried out at the Jet Propulsion Laboratory, Caltech, under a contract with the National Aeronautics and Space Administration.

\subsection*{Supplementary Materials}

\hspace{0.6cm}Materials and Methods

Figures S1-S8

References ({\it 34-49})

\newpage

\begin{figure}[h]
\begin{center}
\includegraphics[width=16cm]{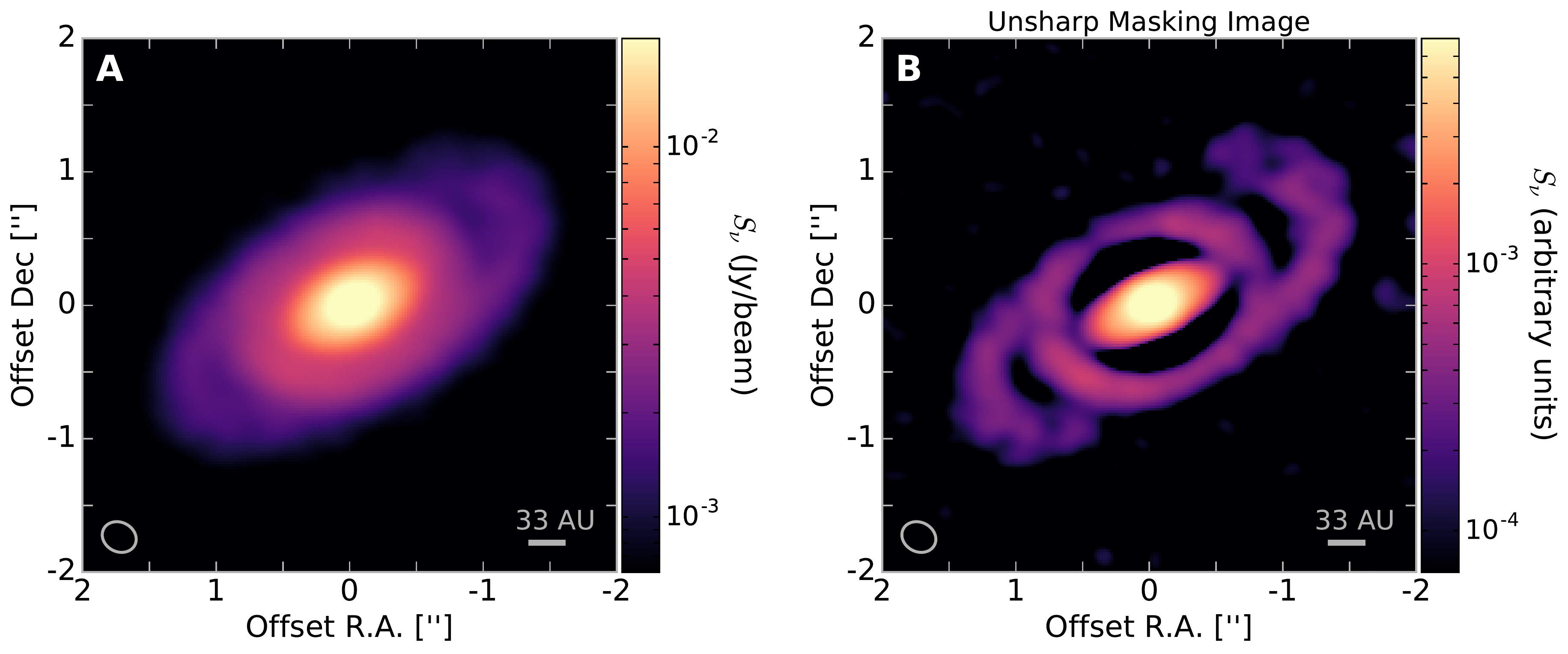}
\end{center}
\end{figure}

\noindent {\bf Fig.\ 1. Thermal dust emission from the protoplanetary disk surrounding Elias 2-27.} The disk was imaged at a wavelength of 1.3 mm with ALMA reaching an angular resolution of $0.26''\times0.22''$ (indicated by the ellipse in the bottom-left corner), which corresponds to $36\times31$~AU at the distance of the star (where AU is the astronomical unit). The field of view center (at~0,0) corresponds to the disk emission peak located at Right Ascension (J2000) = 16h 26m 45.024s, Declination (J2000) = --24d 23m 08.250s and coincidental with the position of the star Elias 2-27. {\bf (A)} 1.3 mm dust continuum image from the Elias 2-27 protoplanetary disk over a $4''\times4''$ area. The color-scale represents flux density measured in units of Jansky per beam (1~$\text{Jy} = 10^{-26}$ W m$^{-2}$ Hz$^{-1}$). {\bf (B)} Increased contrast image from processing the original ALMA observations shown in panel (A) with an unsharp masking filter ({\it 17}). 

\newpage

\begin{figure}[h]
\begin{center}
\includegraphics[width=7.8cm]{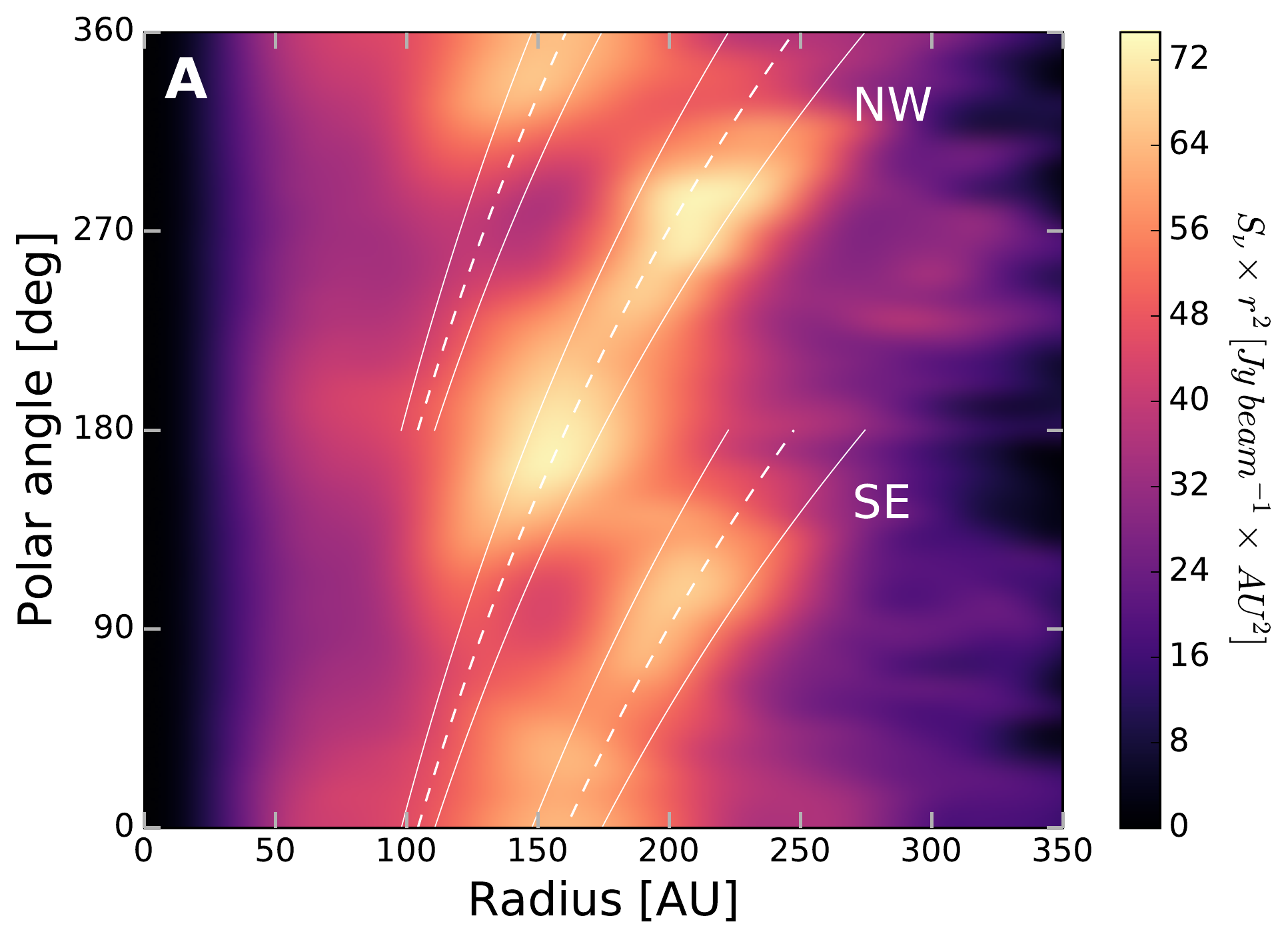}
\includegraphics[width=8.0cm]{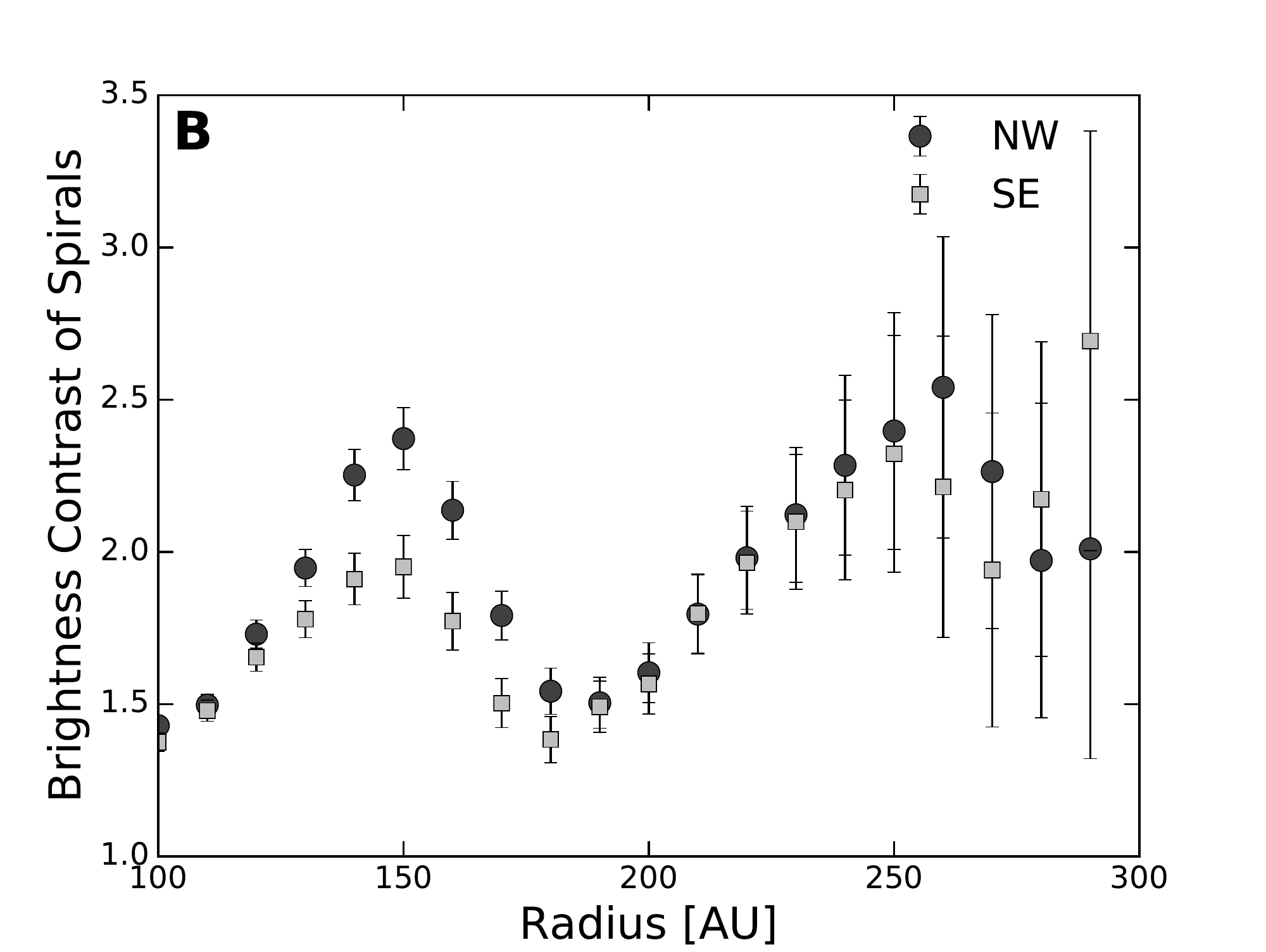}
\end{center}
\end{figure}
\noindent {\bf Fig.\ 2. Polar projection of disk emission and measured contrast over the spirals in the Elias 2-27 protoplanetary disk. (A)} Projection onto polar coordinates (i.e. polar angle θ vs. deprojected radial distance to the central star $R$) of the dust continuum observations from the Elias 2-27 disk. The emission has been scaled by $R^2$ to aid visualization and the polar angle is defined as $\theta=0^{\circ}$ (North) increasing towards East. Curves correspond to the best-fit model spirals for the NW and SE arms (dashed lines) and their constraint at the $3\sigma$ level (solid lines). {\bf (B)} Surface brightness contrast of the continuum emission along each spiral arm, defined as the ratio between the peak of emission and the background surface brightness ({\it 17}), which is computed at increasing radial distance from the star.  

\newpage

\begin{figure}[h]
\begin{center}
\includegraphics[width=7.7cm]{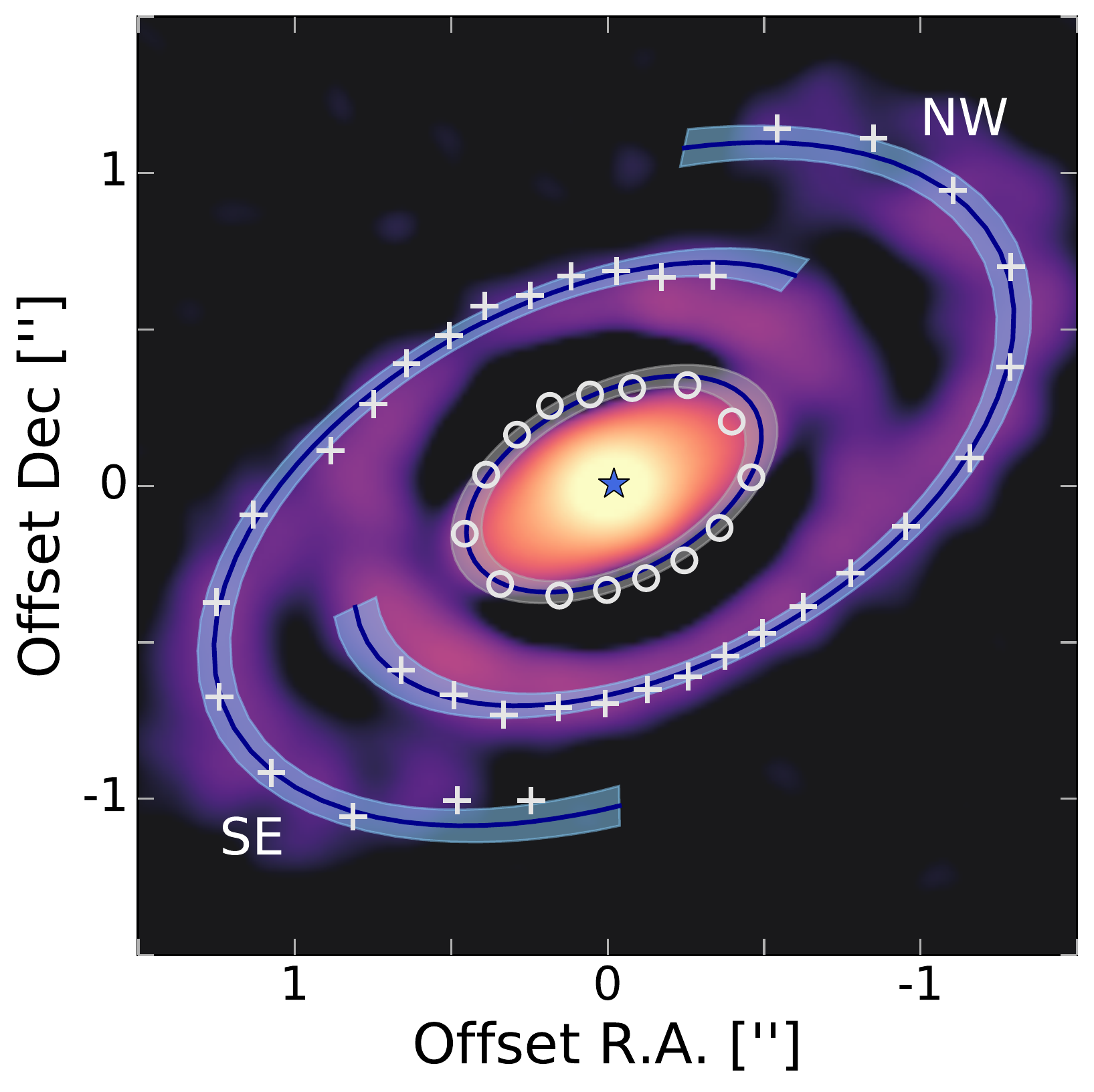}
\end{center}
\end{figure}
\noindent {\bf Fig.\ 3. Model of symmetric spirals and dark ring for the Elias 2-27 protoplanetary disk.} The local maxima (crosses) and local minima (circles) in the continuum emission of Elias 2-27 are indicated. The maxima trace the NW and SE spirals whilst the minima trace a ring. A model with symmetric spirals, that shares the same geometry (inclination and position angle) with the inner dark ring, was able to reproduce the location of the local maxima and minima of emission, as illustrated by the best-fit model (solid curve) and $3\sigma$ constraints (shaded regions). The blue star denotes the position of Elias 2-27. To illustrate the location of these features in the image, we overlaid these results over the unsharp masked image from Fig.~1B, however, this image was not used for the calculations.

\newpage

\subsection*{Materials and Methods}

\noindent \underline{ALMA Observations}

Elias 2-27 was observed as part of the ALMA Cycle 2 program 2013.1.00498.S (P.I.\ L.\ P\'erez). The target was selected based on Disks@EVLA observations (see below) that show emission from cm-sized grains inside 50 AU in the disk, which are indicative of radial segregation by particle size ({\it 34,35}). ALMA observations were obtained with 44 antennas whose baseline lengths ranged between 15.1 m to 1.6 km. The total time on source was 12.5 min with a total continuum bandwidth of 6.8 GHz. The spectral setup consisted of two 2 GHz and three 468.75 MHz windows for continuum observations, and three 468.75 MHz windows centered at 220.183, 219.715, and 230.714 GHz to simultaneously observe the molecular line emission from C$^{18}$O, $^{13}$CO, and $^{12}$CO, respectively, each in the $J = 2-1$ transition. The ALMA pipeline ({\it 36}) and the CASA software ({\it 37}) were used for calibration, and a set of standard calibrators was observed to calibrate the complex interferometric visibilities (PKS J1517-2422 for bandpass, Titan for absolute flux density calibration, and PMN J1626-2426 for gain calibration). 

\vspace{0.2cm}\noindent \underline{ALMA Continuum Imaging}

Amplitude and phase self-calibration was performed on the continuum observations of Elias 2-27 using the multiscale cleaning algorithm ({\it 38}) in CASA, to encompass both the extended and compact emission in the disk. This procedure improved the dynamic range of the dust continuum image from a signal-to-noise ratio (SNR) of 340 to 547. The final image (Fig.\ 1A) was produced with Briggs weighting with a robust parameter of 1.0 that provided the optimal combination of sensitivity and spatial resolution ($0.26''\times0.22''$), reaching an image RMS noise level (hereafter referred to as $\sigma$) of $\sigma = 58.6 $~$\mu$Jy/beam.

\vspace{0.2cm}\noindent \underline{Unsharp Masking Filtering}

We performed unsharp masking ({\it 39}) by subtracting a scaled and smooth version of the original image to remove the large-scale disk emission. This increases the contrast of the observations and highlights the spiral-like structure of the disk in the dust continuum emission. The original image was smoothed with a 2-D Gaussian of $0.33''$ FWHM and the smoothed image was scaled by a factor of 0.87 before subtracting it from the original. Because the disk is inclined with respect to our line-of-sight, the spiral arms appear circular at small radii (closer to the gap) while they seem more elliptical farther out. The unsharp masking technique accentuates the spirals while removing the overall large-scale disk emission (filtering the disk emission up to a certain spatial scale, given the fixed circular Gaussian used). Thus, the central core is not completely removed, the spiral features are highlighted, and a wide cavity appears. This cavity is wider and deeper along the disk minor-axis than along its major axis because the major-axis does not suffer from inclination effects. 

\newpage
\vspace{0.2cm}\noindent \underline{Radiative Transfer Model}

We performed radiative transfer calculations using RADMC3D ({\it 26}) in order to compute the dust temperature and surface density, as well as the optical depth of the dust continuum emission at 1.3 mm as a function of radial distance from the star ($R$). For simplicity, we adopt a previous surface density constraint found for this object ({\it 20}), which assumes axial symmetry. Given that the contrast of the observed features (spirals and gap) is small, the azimuthally averaged solution will be representative of the dust temperature, surface density, and optical depth at different radial locations in the disk. 
First, we compute the radial profile of the dust continuum emission by averaging the observed 1.3 mm flux density in radial bins of half of the beam size (data points in Figure S1A). We adopt a surface density profile that corresponds to the similarity solution for a viscously accreting disk in Keplerian rotation around a central object ({\it 40}), which can be described by a power law combined with an exponential taper at large radius: 
\begin{equation}
    \Sigma = \Sigma_c \left(\frac{R}{R_c} \right)^{-\gamma} \exp\left[- \left(\frac{R}{R_c} \right)^{2-\gamma} \right] \tag{E1}\label{myeq}
\end{equation}

In Eq. \eqref{myeq}, $\Sigma_c$ corresponds to the normalization surface density, $R_c$ is the characteristic radius, and $\gamma$ is the surface density gradient. The parameters adopted, based on the previous constraint on this disk ({\it 20}), are: $\Sigma_c = 0.05$ g cm$^{-2}$, $R_c = 200$~AU, $\gamma = 0.7$, with a dust mass opacity $\kappa_{1.3mm} = 2.4$ cm$^2$ g$^{-1}$. This results in a total disk mass of $M_{disk} = 0.1 M_{\odot}$, assuming a standard gas-to-dust-ratio of 100. The results from our radiative transfer modeling, including the midplane dust temperature, surface density, and optical depth profile, are found in Figure~S1. 

\vspace{0.2cm}\noindent \underline{Calculation of Spiral Arms Contrast}

We define the contrast of the spiral arms as $S_{\nu}^{\text{Spiral}}(R) / S_{\nu}^{\text{Bg}}(R)$ where $R$ is the deprojected radial distance from the star in the ALMA 1.3 mm image, $S_{\nu}^{\text{Spiral}}(R)$ is the peak surface brightness of the spiral feature at radius $R$, and $S_{\nu}^{\text{Bg}}(R)$ is the background surface brightness on the disk, defined as the minimum flux density at radius $R$.

\vspace{0.2cm}\noindent \underline{Estimation of Toomre $Q$ parameter and Dust Trapping in the GI scenario}

From the derived temperature profile and surface density profile in our radiative transfer calculation, we derive the Toomre $Q$ instability parameter ({\it 23}). For a dust mass opacity of $\kappa_{1.3mm} = 2.3$ cm$^2$ g$^{-1}$, commonly used in the literature ({\it 41}), and for a ratio between the gas mass and dust mass in the disk consistent with the interstellar medium value of 100 ({\it 42}), we obtain the Toomre $Q$ values illustrated by the solid curve of Figure S2. The minimum value of the Toomre $Q$ parameter is well inside the stable regime and occurs at a radius of about 150~AU. The absolute value of $\kappa_{1.3mm}$ and the gas-to-dust ratio are poorly constrained quantities in protoplanetary disks. Considering a factor of 4 lower dust mass opacity than the canonical value used above, we calculate the value of Toomre $Q$ to be marginally gravitationally unstable right at the radii where we find the spiral arms (Fig.\ S2, dashed curve).
If the disk were gravitationally unstable, the observed spirals in the dust continuum would arise due to enhancements in gas density where particles with a Stokes number ($St$) close to unity will be efficiently trapped and concentrated ({\it 43}), creating the spiral structures that we observe in dust continuum emission. For the particles traced by the ALMA observations presented here, i.e. those with size $a \approx 1$~mm, we estimate an Stokes number of $St \approx \pi/2 \times \rho \times a / \Sigma \approx 0.1$ ({\it 43}), assuming a standard physical density for these grains of $\rho \approx 1$ g cm$^{-3}$ and recalling that over the location of the spirals the disk surface density is roughly $\Sigma \approx 1.5$ g cm$^{-2}$ (c.f.~Fig.~S1, and assuming a gas-to-dust ratio of 100). Thus, the dust grains traced by these ALMA observations in the Elias~2-27 disk would only experience weak trapping and concentration if the gravitational instability mechanism were at work.

\vspace{0.2cm}\noindent\underline{Fitting of Dark Ring, Spiral Arms and Large-Scale Disk Geometry}

We fit a smooth and strictly monotonic intensity profile (solid curve, Fig.\ S3A) to the observed radial profile (data points, Fig.\ S3A) and compute a residual profile from subtracting the best-fit to the emission radial profile (Fig.\ S3B). We find an excess of emission between 120 to 250 AU, as well as a lack of emission between $\sim50$-100~AU. We subtract the best-fit axi-symmetric radial profile from the ALMA image and create a residual map where the non-symmetric structure is highlighted. In this residual map we find the position of local maxima and minima of the emission, at evenly spaced azimuthal angles (spaced every $12^{\circ}$). We fit an ellipse to the local minima points and two logarithmic spirals to the local maxima points, with their center position, inclination, and position angle as free parameters as explained in the main text. To find the best-fit and errors of the dark ring and spiral arms fit, illustrated in Figure 3, we employed a Markov Chain Monte Carlo method ({\it 44}). The best-fit parameters and results of this fitting are presented in the main text and Figure 3.
To constrain the large-scale disk geometry we fit a Gaussian function to visibilities on baselines smaller than 150 m; at this resolution the spiral structure and the dark ring are not resolved, while the large-scale disk is. We find excellent agreement between the orientation on the sky of the large-scale disk ($i_{disk} = 54.4^{\circ} \pm 1.4^{\circ}$, $PA_{disk} = 118.8^{\circ} \pm 1.5^{\circ}$) and that of the spiral arms and dark ring, which were fitted separately as explained above.

\vspace{0.2cm}\noindent\underline{ALMA Spectral Line Observations of $^{12}$CO, $^{13}$CO, and C$^{18}$O Molecules}

The amplitude and phase gain solutions found from self-calibration of the continuum observations were applied to the spectral line data of the $J = 2-1$ rotational transition of $^{12}$CO, $^{13}$CO, and C$^{18}$O. We adopted a natural weighting scheme (robust parameter of 2.0) for best sensitivity, and as in the dust continuum observations, we used the multiscale cleaning algorithm in CASA to deconvolve the image. To reduce the level of contamination from the surrounding molecular cloud, baselines shorter than 20 m, corresponding to spatial scales larger than $16.4''$ or 2280 AU, were excluded. The uv-coverage of these observations reveals that a 20 m threshold only removes a couple of baselines; doubling or tripling this threshold did not significantly improve the problem of cloud contamination/absorption while it removed many more baselines.

The $^{12}$CO observations have a resolution of $0.27''\times0.23''$ (Fig.\ S5); for the $^{13}$CO and C$^{18}$O observations (Fig.\ S6 and S4) we applied uv-tapering of the visibilities to down-weight long baselines, resulting in lower resolution images of $0.42''\times0.38''$ angular resolution with improved SNR. The sensitivity achieved on each image is indicated on the respective figure caption. The spectral line observations contain emission brighter than 3 times the noise level~($3\sigma$) from $v_{\text{LSR}} = -3.6$ km s$^{-1}$ to $+8.1$ km s$^{-1}$ for $^{12}$CO, from $v_{\text{LSR}} = -1.3$ km s$^{-1}$ to $+6.0$ km s$^{-1}$ for $^{13}$CO, and from $v_{\text{LSR}} -1.1$ km s$^{-1}$ to $+4.9$ km s$^{-1}$ for C$^{18}$O. The spatially resolved line emission from $^{12}$CO traces optically thick gas and is detected out to a 630 AU radius, while the emission from the less abundant $^{13}$CO and C$^{18}$O isotopologues, which will have lower optical depth, is detected out to a 630 and 500 AU radius, respectively. The large spatial extent in the gas lines, compared to the more compact disk observed in the dust emission, has been seen in other disks as well ({\it 45}). Note that the $^{12}$CO and isotopologues emission extends beyond where the expected freeze out temperature of $^{12}$CO should be reached ($T_{\text{freeze-out}}\sim 20$~K), likely due to the known disk flaring ({\it 20}) that warms up the surface layer of the disk that traces most of the $^{12}$CO emission, consistent with the brightness temperature of $^{12}$CO at large disk radii to be above 20~K when the emission is above $10\sigma$.

Significant absorption and resolved-out emission from the surrounding molecular cloud is apparent in both $^{12}$CO and $^{13}$CO molecules, particularly at $v_{\text{LSR}}= 2$ to 6 km s$^{-1}$, while the C$^{18}$O observations (Fig.\ S6) do not suffer as severely from molecular cloud contamination. From these data we identify the systemic velocity of the star to be between $v_{\text{LSR}} \approx 2.0$ to $2.2$~km~s$^{-1}$, offset from the $\rho$-Ophiuchus systemic velocity of order 5 km s$^{-1}$ ({\it 46}). We generate a position-velocity diagram (Fig.\ S7) using the spectral line observations from $^{12}$CO, as these data possess the highest SNR and the best velocity-resolution of the three molecules observed. A lack of emission is observed at $v_{\text{LSR}} > 2$ km s$^{-1}$ owing to the disk emission being heavily absorbed by the $\rho$-Ophiuchus cloud. We compute the corresponding Keplerian profile for a stellar mass of $0.5 \pm 0.2\; M_{\odot}$ and a systemic velocity of $2.1 \pm 0.1$ km s$^{-1}$ (including the uncertainties in mass and systemic velocity for Elias 2-27), and find no significant deviations from a Keplerian velocity pattern in the Elias 2-27 disk.

\vspace{0.2cm}\noindent\underline{Absolute Geometry of the System from Optically Thick $^{12}$CO Observations}

Near the systemic velocity of the star (at $v_{\text{LSR}} \approx 2.0$-2.2 km s$^{-1}$) we observe that of the two lobes that should be aligned with the disk minor axis (dashed line in Figure S5), the NE lobe is brighter and more compact in $^{12}$CO than in the SW (there is no evidence for an outflow in Elias 2-27, so ``lobe'' is used here to signify the disk emitting region near the systemic velocity of the star rather than an outflow). In the panel inset of Figure S5, which zooms into a $1.25''\times1.25''$ area centered on the star (marked by a black circle), we measure a brightness asymmetry between the NE lobe and the SW region to be larger than $3\sigma$, at an offset of $0.5''$ along the minor axis of the disk.

Since the disk is inclined with respect to our line of sight and since the $^{12}$CO$(J = 2-1)$ line is optically thick, we expect that these observations trace a surface that has a non-negligible vertical extent. For an inclined disk, the projection of that curvature will make the lobe on the far side of the disk brighter and more compact, as demonstrated in other ALMA protoplanetary disk observations (e.g.~{\it 45}). Assuming the NE/SW brightness asymmetry near the systemic velocity and at $0.5''$ from the star along the disk minor axis, is due to the high optical depth of the lines, we infer that the SW side of the disk is tilted towards Earth. Since the rotation sense of the disk is also determined directly from the resolved spectral line observations, and since for Elias 2-27 the disk rotates from the NW to the SE (see Fig.\ S4, S5, S6), we conclude that the observed spirals NW and SE point away from the direction of rotation, i.e. these are trailing spiral arms.

\vspace{0.2cm}\noindent\underline{Karl G. Jansky Very Large Array (VLA) Observations of Elias 2-27 and the Possibility of Grain}

\vspace{0.0cm}\noindent\underline{Growth in the 70 AU Gap}

Elias 2-27 was observed as part of our Disks@EVLA program (project code AC982, P.I. C. Chandler) at 9 mm with the Ka-band receiver of the VLA between February 2011 and October 2012, providing baseline lengths between 60 m to 36 km. Further details on calibration and imaging are presented in our group’s publications ({\it 34,35,47,48}) while a brief summary is presented here. We employed dual-polarization receivers with a total continuum bandwidth of 2 GHz and two independently tunable basebands at 30.5 and 37.5 GHz. We calibrated the complex interferometric visibilities from observations of a set of standard calibrators (3C 279 for bandpass calibration, 3C 286 for absolute flux density calibration, and NVSS J162546-252739 for gain calibration). The VLA pipeline ({\it 49}) and the CASA software ({\it 36}) were used for calibration and imaging. 

Figure S8 presents the distribution of dust continuum emission at 9 mm from Elias 2-27, overlaid to the ALMA observations of Figure 1. These observations reach an angular resolution of $0.15''\times0.11''$ ($21\times14$~AU). No emission at 9 mm is detected outside of a radius of about 50~AU, indicative of the presence of large particles only out to that disk radius. Thus, the gap at 70 AU likely traces a decrement in density of material rather than a pure grain growth effect.

\newpage
\begin{figure}[h]
\begin{center}
\includegraphics[width=7.9cm]{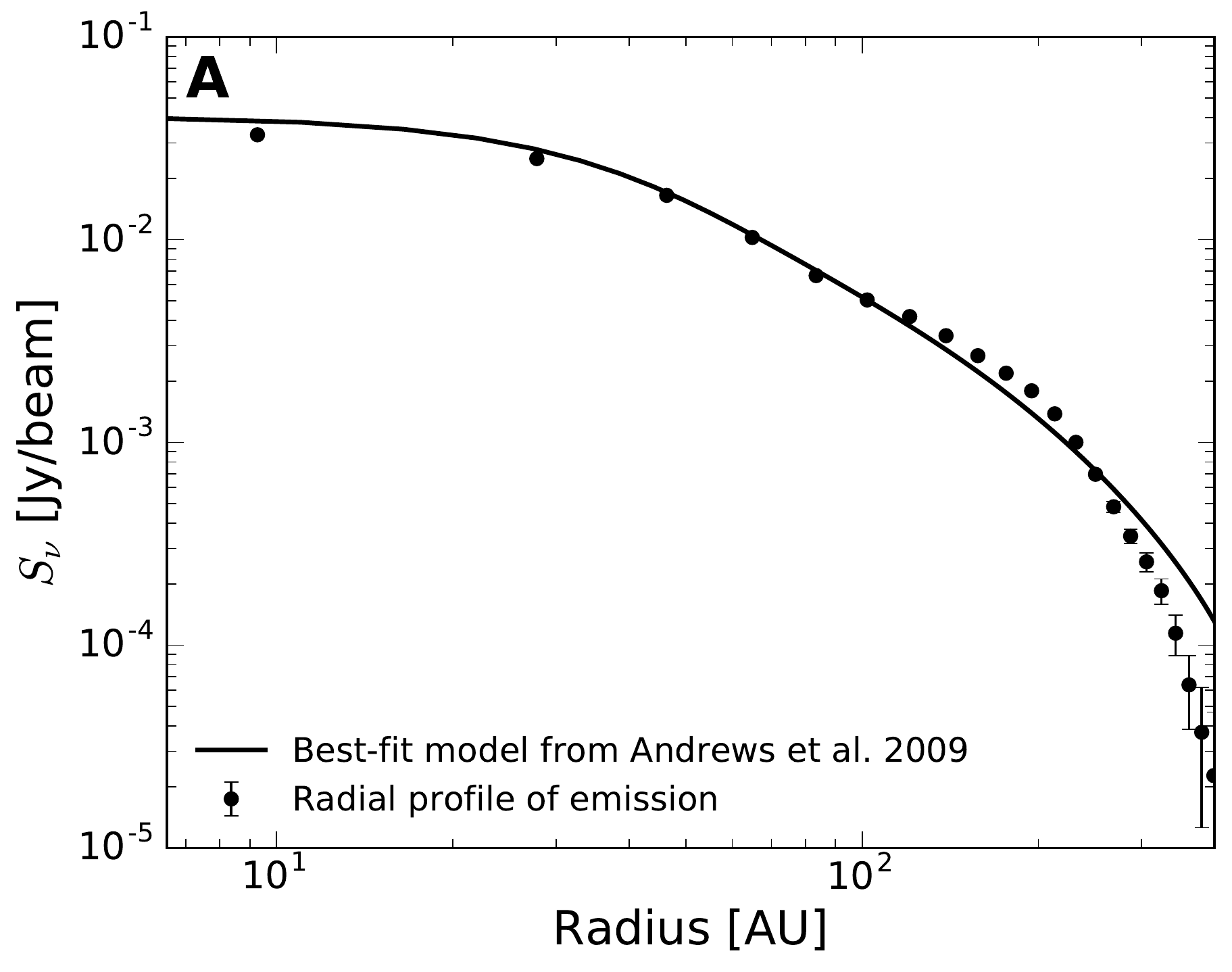}
\includegraphics[width=7.9cm]{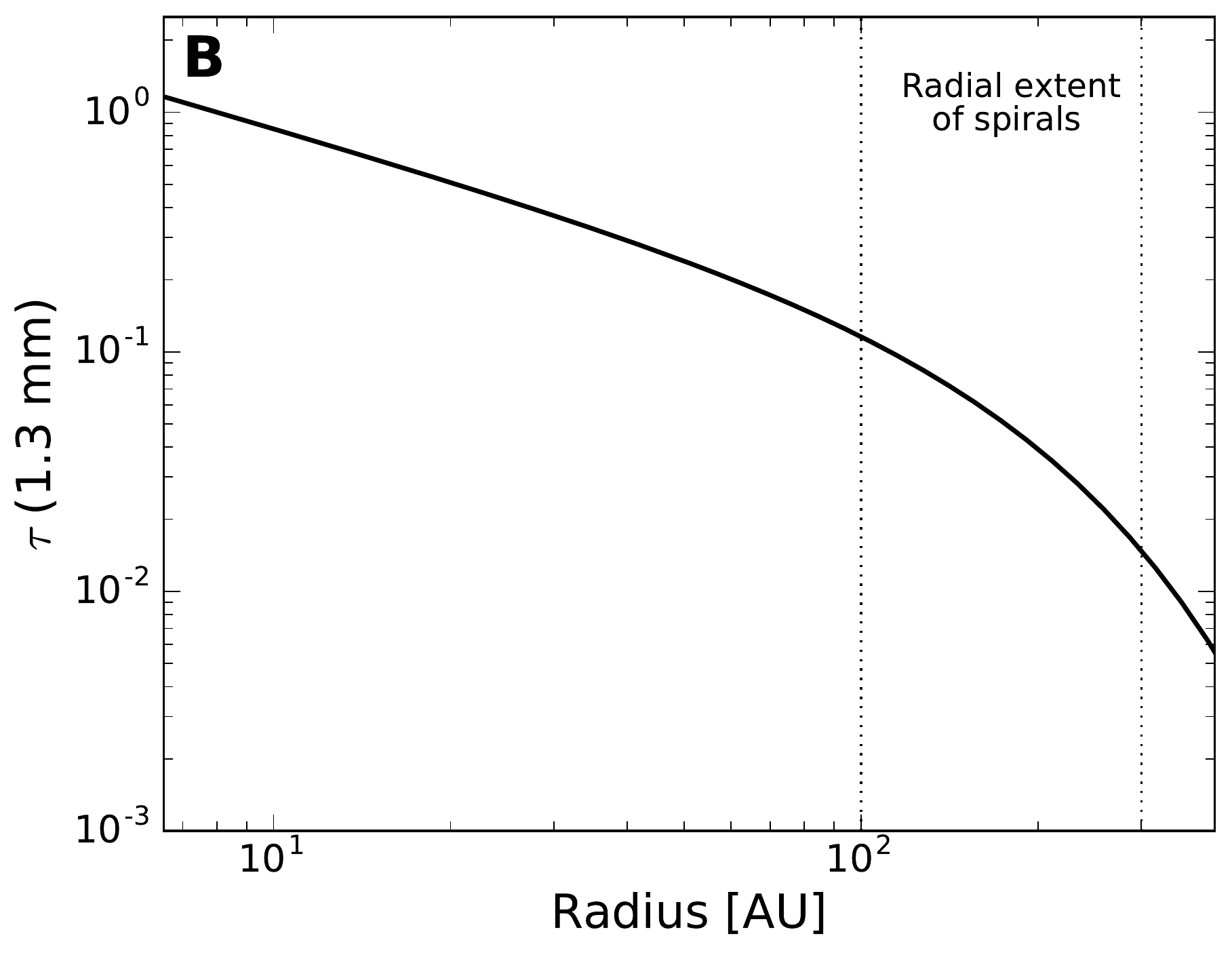}
\includegraphics[width=7.9cm]{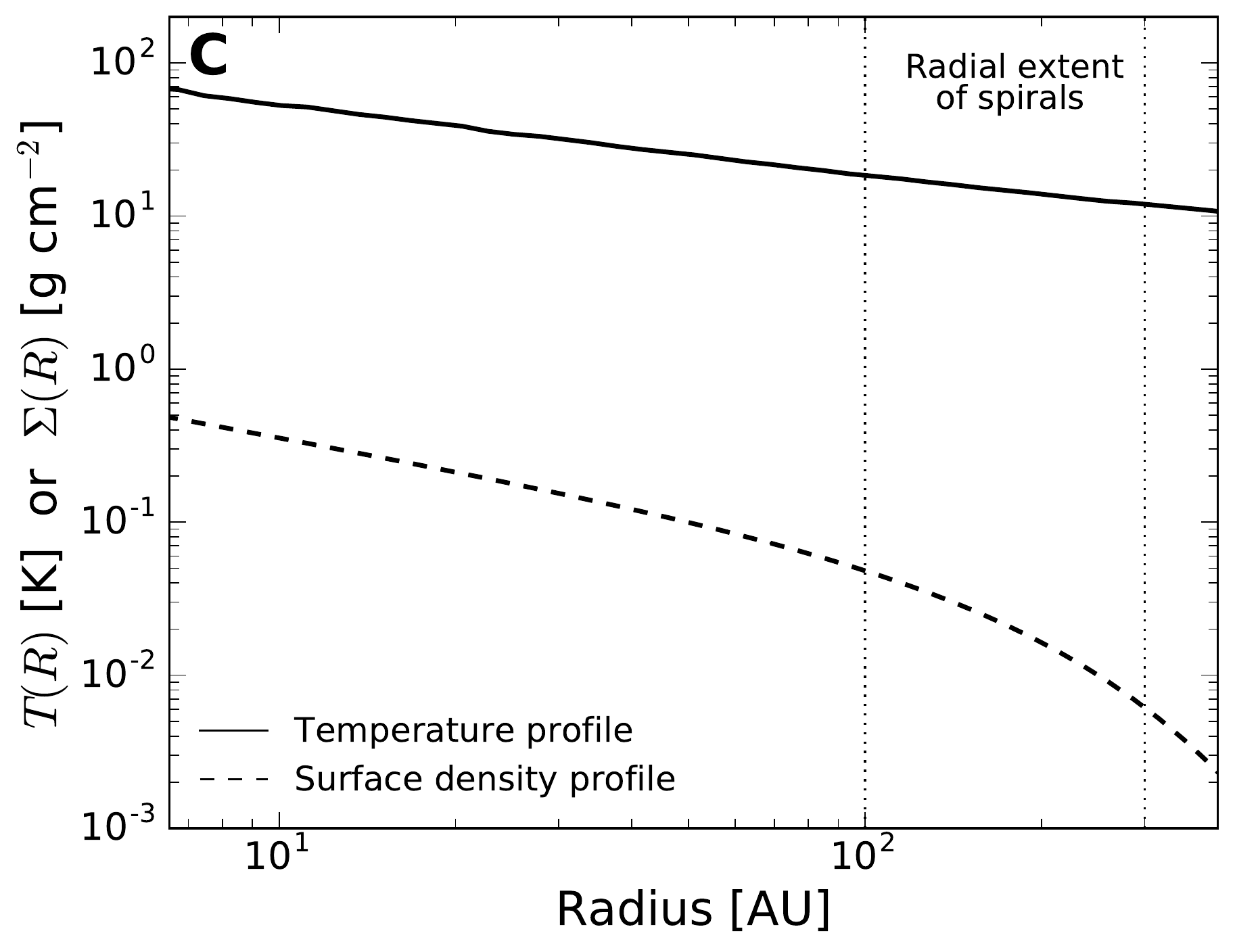}
\end{center}
\end{figure}
\noindent {\bf Fig.\ S1. Results from radiative transfer on the Elias 2-27 disk emission profile at 1.3 mm. 
(A)} The azimuthally averaged radial profile of the disk emission (points) and the previously obtained best-fit model ({\it 20}), which we select as representative of the disk emission at this wavelength. {\bf (B)} The inferred optical depth of the emission at 1.3 mm from this representative radiative transfer model. {\bf (C)} The dust temperature at the disk midplane and the surface density profiles in Elias 2-27 derived from this representative radiative transfer model.

\newpage

\begin{figure}[h]
\begin{center}
\includegraphics[width=9.5cm]{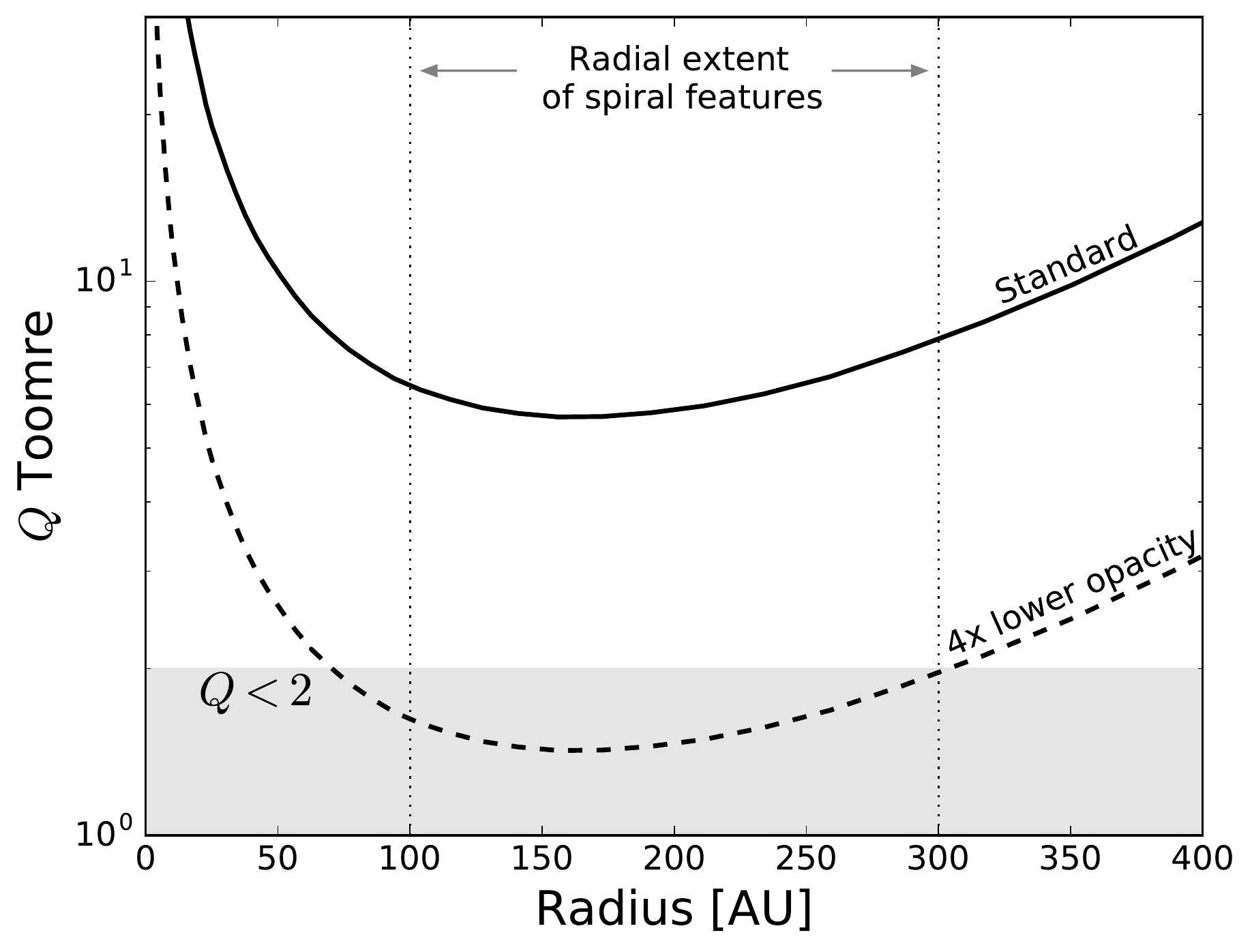}
\end{center}
\end{figure}
\vspace{-.7cm}
\noindent {\bf Fig.\ S2. Constraints on the Toomre $Q$ instability parameter throughout the Elias 2-27 disk.}
Two different values for the dust opacity are assumed (a ``standard''' value of $\kappa_{1.3mm} = 2.3$ cm$^2$ g$^{-1}$, and one 4 times lower) while the gas-to-dust ratio is kept fixed at 100. The shaded region corresponds to the unstable regime ($Q \le 2$); the radial extent of the spiral arms is marked with two vertical lines.

\begin{figure}[h]
\begin{center}
\includegraphics[width=9.5cm]{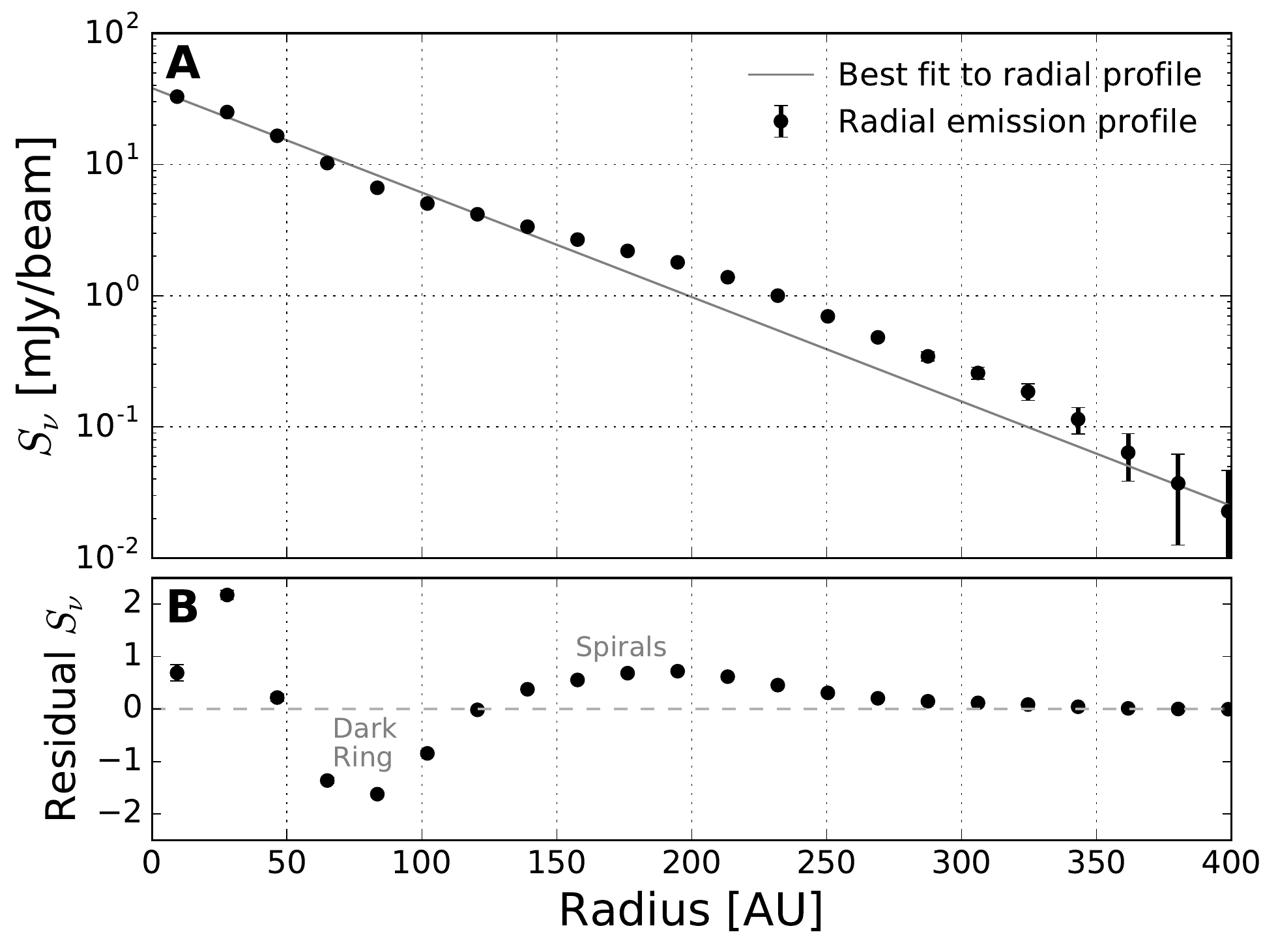}
\end{center}
\end{figure}
\vspace{-.7cm}
\noindent {\bf Fig.\ S3. Azimuthally-averaged radial profile of emission and residuals. (A)} Radial profile of the emission at 1.3 mm (data points) and best-fit intensity profile (solid line). {\bf (B)} Residual from subtracting the best-fit intensity profile to the radial emission profile, where two regions are highlighted: lack of emission at about 70 AU, excess of emission between 120-250 AU.

\newpage
\begin{figure}[h]
\begin{center}
\includegraphics[width=12.8cm]{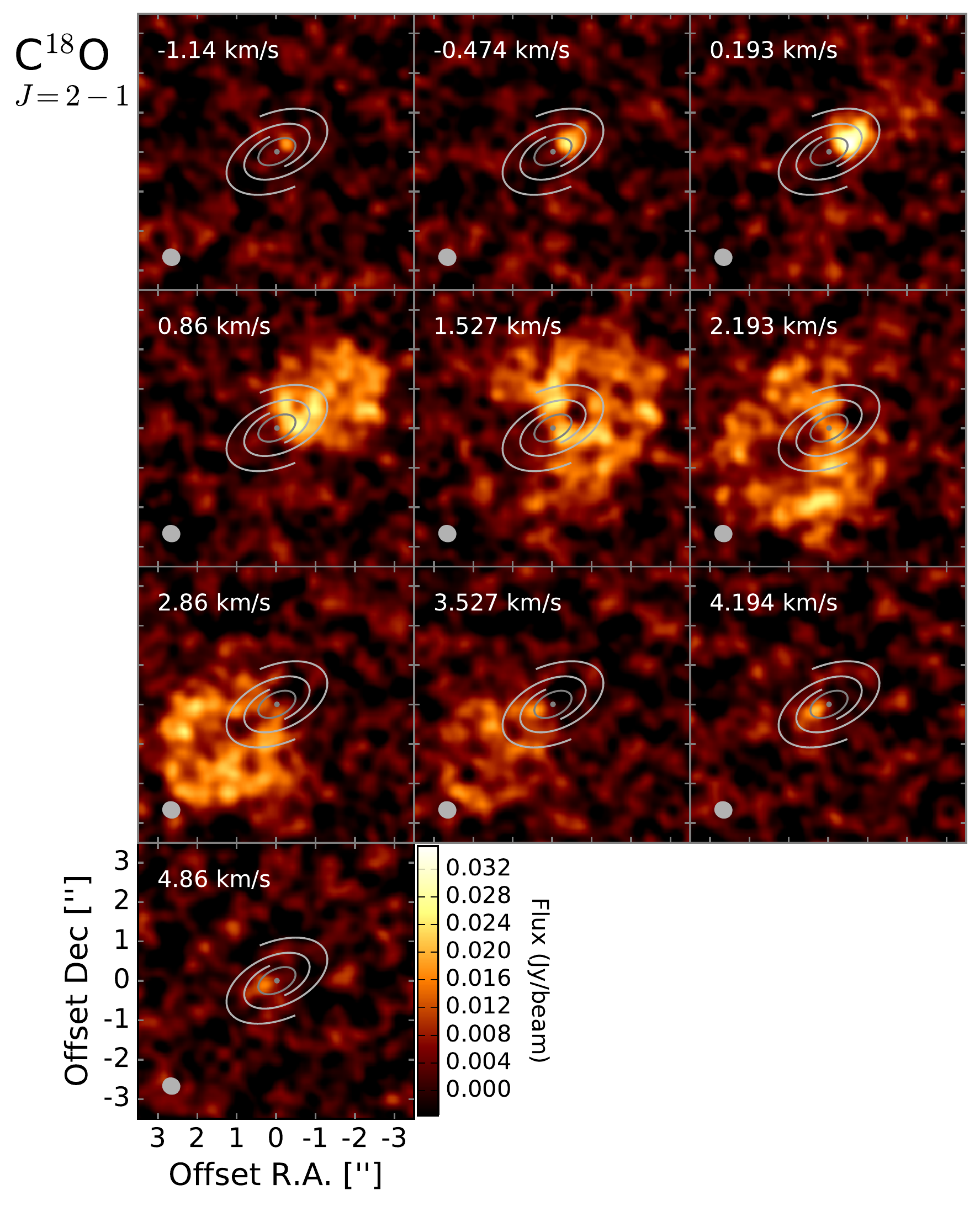}
\end{center}
\end{figure}
\noindent {\bf Fig.\ S4. Velocity channel maps of the $J = 2-1$ rotational transition of C$^{18}$O in Elias 2-27.}
Each panel corresponds to a different $v_{\text{LSR}}$ velocity observed, reaching a sensitivity of $\sigma = 3.6$~mJy/beam in a 0.67 km s$^{-1}$ channel. The resolution ($0.42''\times0.38''$, $58\times53$ AU) is indicated by the ellipse in the bottom-left corner. The field of view corresponds to a $7''\times7''$ area whose center is the same as Figure 1. The best-fit dark ring and spiral arms (white lines) illustrate the extent of the dust emission in the disk.

\newpage
\begin{figure}[h]
\begin{center}
\vspace{-.7cm}
\includegraphics[width=12.7cm]{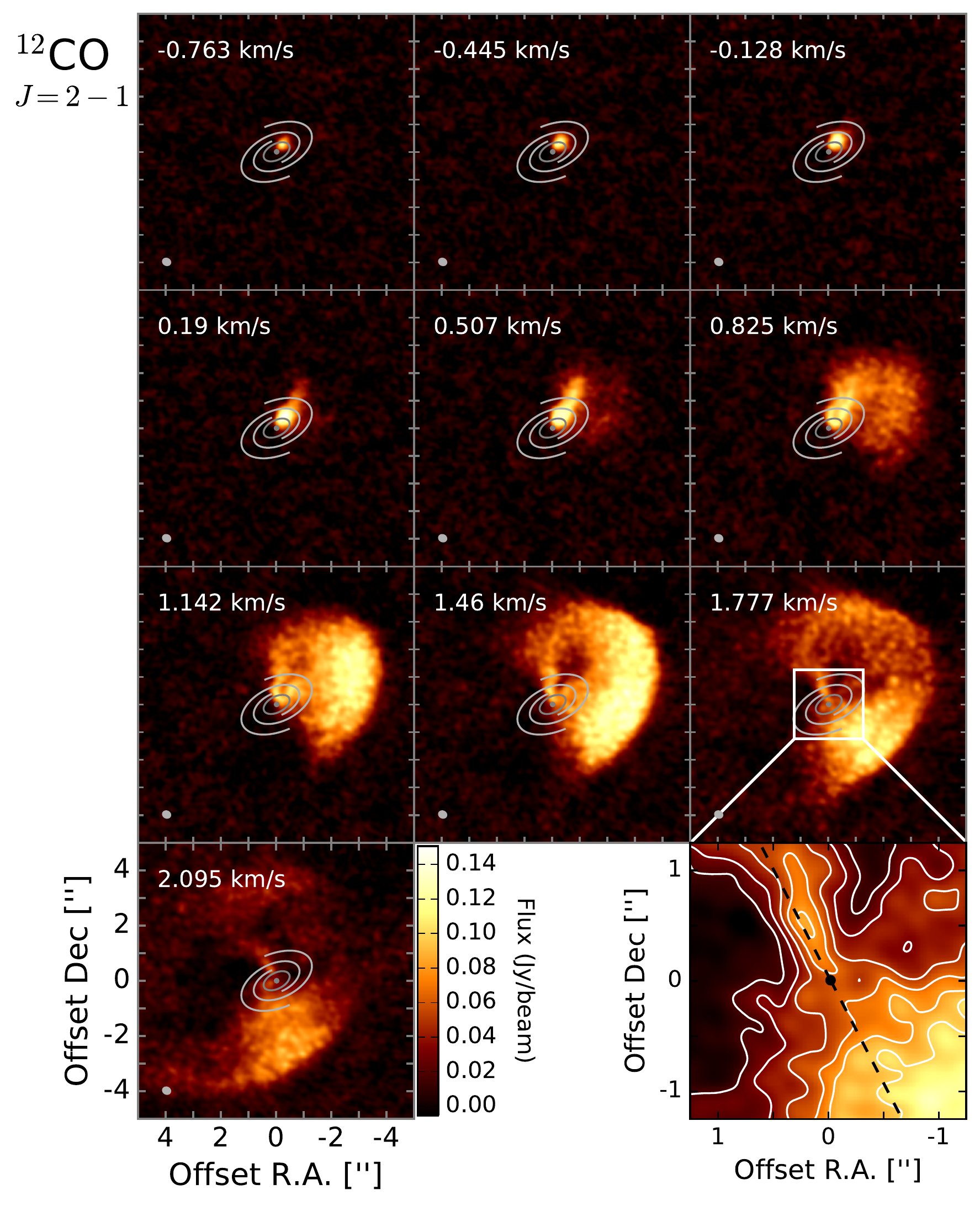}
\end{center}
\end{figure}
\vspace{-0.7cm}
\noindent {\bf Fig.\ S5. Velocity channel maps of the $J = 2-1$ rotational transition of $^{12}$CO in Elias 2-27.}
Each panel corresponds to a different $v_{\text{LSR}}$ velocity observed, reaching a sensitivity of $\sigma = 5.5$~mJy/beam in a 0.32 km s$^{-1}$ channel (only channels at $v_{\text{LSR}} \le 2.0$~km~s$^{-1}$ are shown, these have minimal contamination from the surrounding molecular cloud). The resolution ($0.27''\times0.23''$, $38\times32$~AU) is indicated by the ellipse in the bottom-left corner. The field of view corresponds to a $10''\times10''$ area whose center is the same as Figure 1. The best-fit dark ring and spiral arms (white lines) illustrate the extent of the dust emission in the disk. The panel inset zooms to a $1.25''\times1.25''$ area centered on the star (black circle). Here contours are separated by $3\sigma$ to illustrate the brightness asymmetry along the minor axis of the disk (dashed line), where at an offset of $0.5''$, the NE lobe is brighter than the SW at more than $3\sigma$.

\newpage
\begin{figure}[h]
\begin{center}
\includegraphics[width=12.8cm]{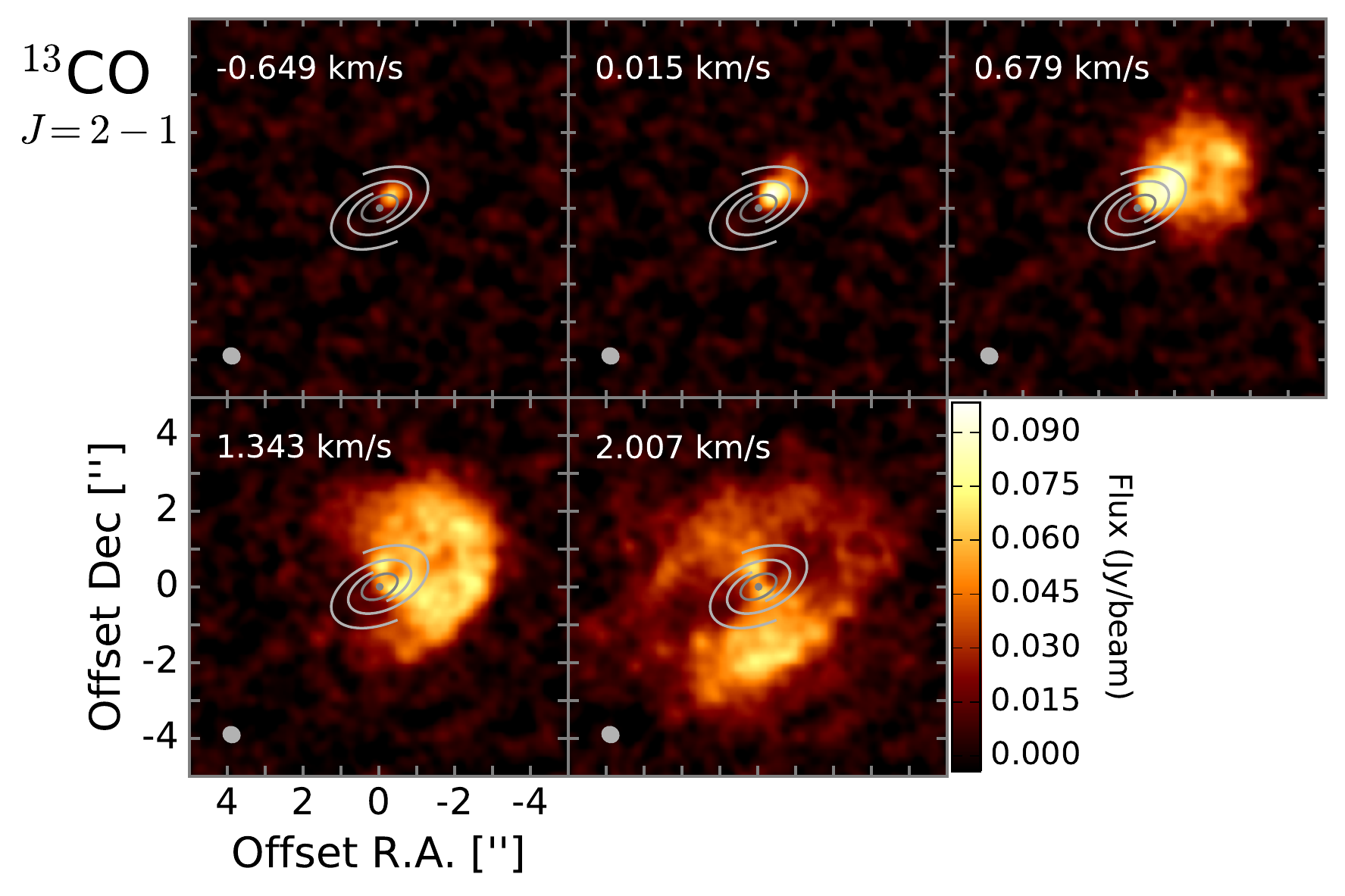}
\end{center}
\end{figure}
\noindent {\bf Fig.\ S6. Velocity channel maps of the $J = 2-1$ rotational transition of $^{13}$CO in Elias 2-27.} Each panel corresponds to a different $v_{\text{LSR}}$ velocity observed, reaching a sensitivity of $\sigma = 4.4$~mJy/beam in a 0.66 km s$^{-1}$ channel (only channels at $v_{\text{LSR}} \le 2.0$~km~s$^{-1}$ are shown, these have minimal contamination from the surrounding molecular cloud). The resolution ($0.42''\times0.38''$, $58\times53$~AU) is indicated by the ellipse in the bottom-left corner. The field of view corresponds to a $10''\times10''$ area whose center is the same as Figure 1. The best-fit dark ring and spiral arms (white lines) illustrate the extent of the dust emission in the disk.

\newpage

\begin{figure}[h]
\begin{center}
\includegraphics[width=9cm]{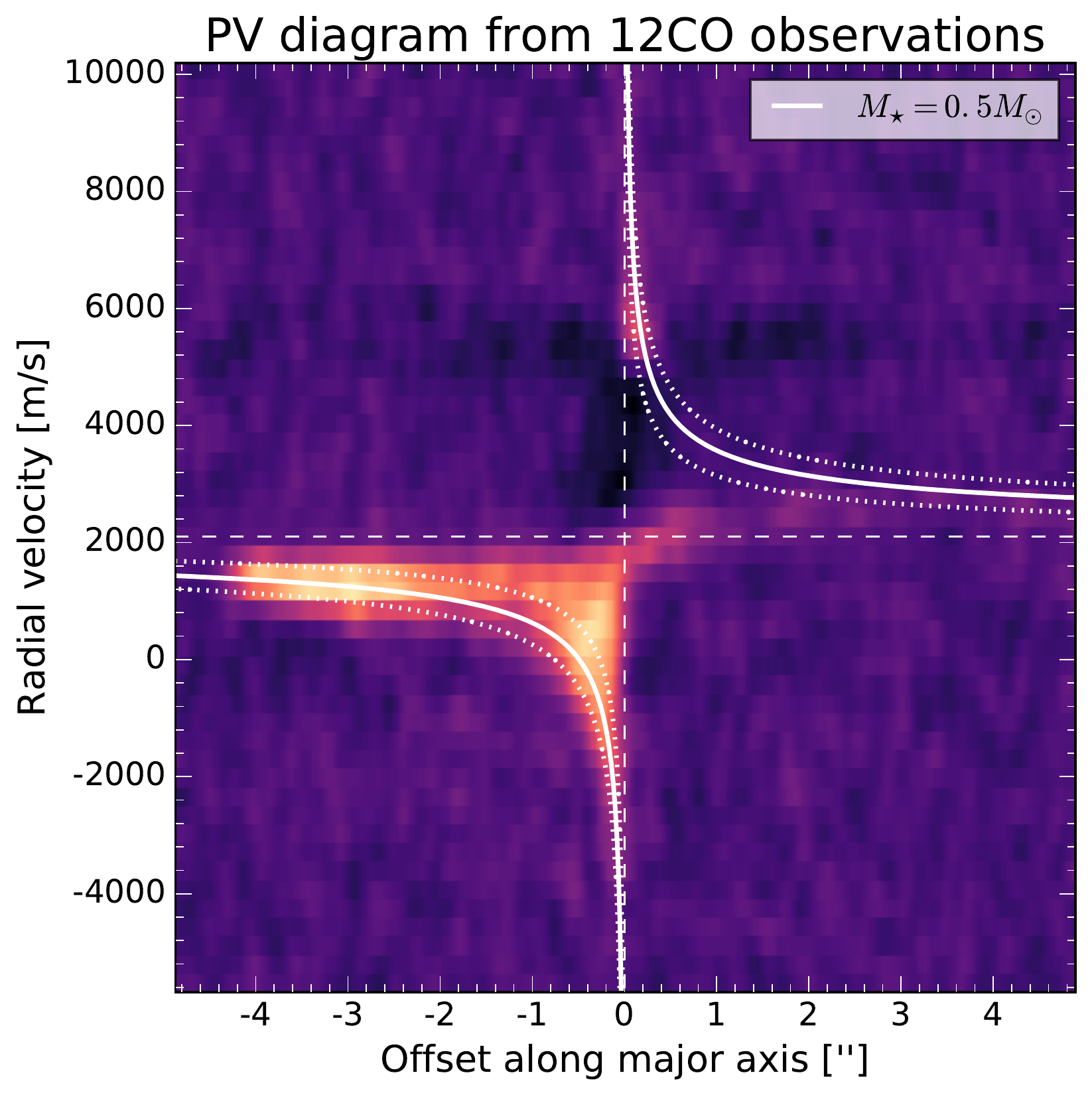}
\end{center}
\end{figure}
\noindent {\bf Fig.\ S7. Position-Velocity diagram of the $J = 2-1$ rotational transition of $^{12}$CO in Elias 2-27.}
We compute the emission from $^{12}$CO along the major axis of the disk (at a $PA = 118.8^{\circ}$) over the entire velocity range of the observations. The $x$-axis corresponds to the positional offset along the disk major axis, while the $y$-axis indicates the $v_{\text{LSR}}$ velocity of the emission. For $v_{\text{LSR}} \le 2.0$~km~s$^{-1}$ little emission is detected, due to absorption by the surrounding molecular cloud. The corresponding Keplerian profiles for a stellar mass of $0.5 \pm 0.2 M_{\odot}$ and a systemic velocity of $2.1 \pm 0.1$~km~s$^{-1}$ is indicated by the white curves.

\newpage
\begin{figure}[h]
\begin{center}
\includegraphics[width=10cm]{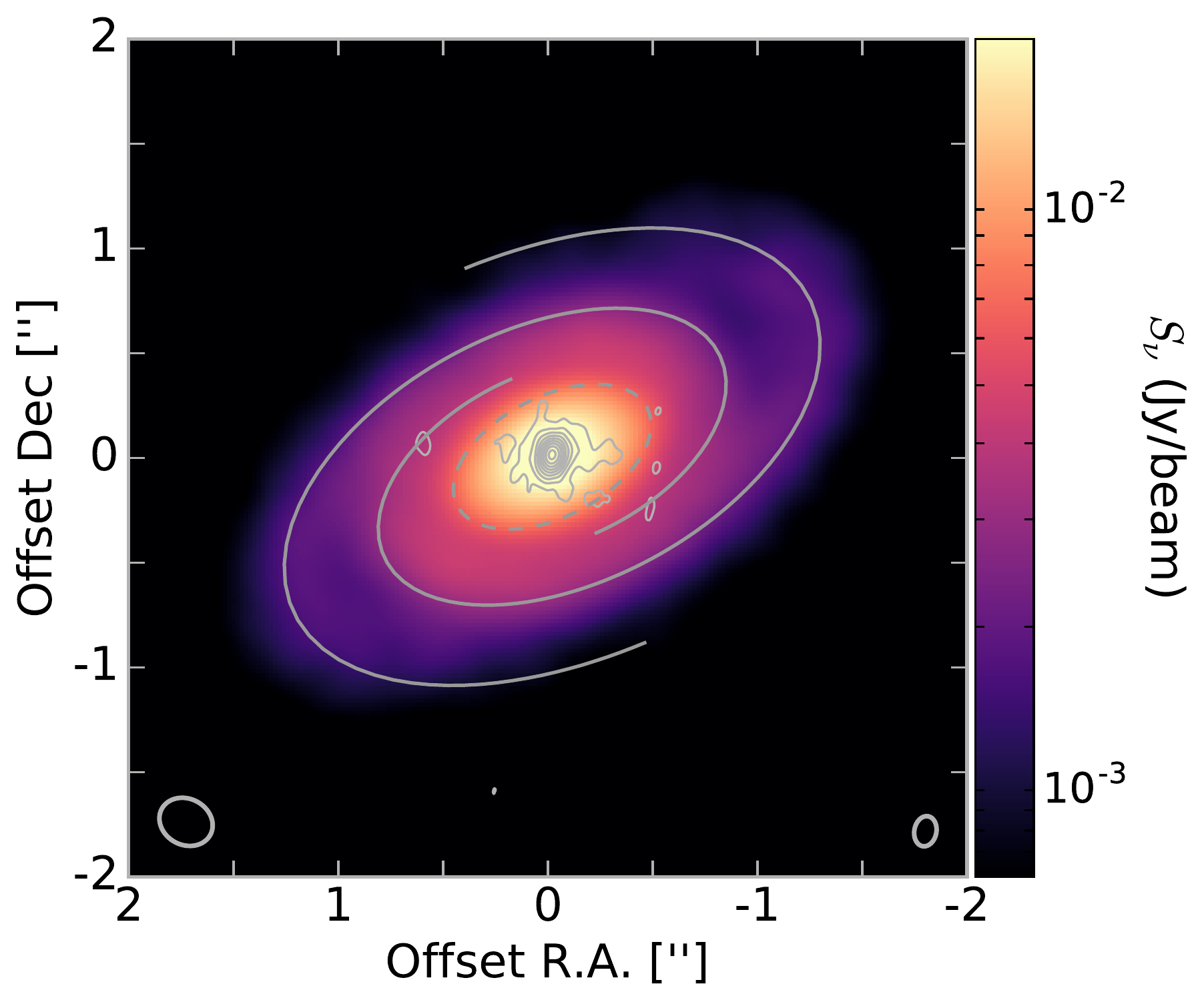}
\end{center}
\end{figure}
\vspace{-0.5cm}
\noindent {\bf Fig.\ S8. VLA observations at 9 mm and ALMA observations at 1.3 mm of the Elias 2-27 disk.}
The dust continuum emission at 9 mm (contours) overlaid to the 1.3 mm ALMA observations of Figure 1. Contours start at $3\sigma$ and are spaced by $3\sigma$ where $\sigma = 0.01$~mJy/beam corresponds to the sensitivity of these VLA observations. Emission at high significance is only detected at $R<50$~AU for the 9 mm observations.

\subsection*{References and Notes}
\begin{enumerate}
  \setcounter{enumi}{33}

\item L. M. P\'erez, et al. Constraints on the radial variation of grain growth in the AS 209 circumstellar disk. Astrophys. J. 760, L17 (2012).

\item L. M. P\'erez, et al. Grain growth in the circumstellar disks of the young stars CY Tau and DoAr 25. Astrophys. J. 813, 41 (2015).

\item H. Shinnaga, et al. ALMA Pipeline: Current Status in Revolution in Astronomy with ALMA: The Third Year, Iono D., Tatematsu K., Wootten A., Testi L., Eds. (Publ. Astron. Soc. Pac. Conference Series, San Francisco, CA, 2015) vol. 499, p. 355.

\item International Consortium of Scientists. CASA: Common Astronomy Software Applications. Astrophysics Source Code Library, record ascl:1107.013 (2011).
\item U. Rau, T. J. Cornwell. A multi-scale multi-frequency deconvolution algorithm for synthesis imaging in radio interferometry. Astron. Astrophys. 532, A71 (2011).
\item D. F. Malin. Unsharp masking. AAS Photo Bulletin. 16, 10-13 (1977).
\item D. Lynden-Bell, J. E. Pringle. The evolution of viscous discs and the origin of the nebular variables. Mon. Not. R. Astron. Soc. 168, 603-607 (1974).
\item S. V.W. Beckwith, A. I. Sargent, R. S. Chini, R. Guesten. A survey for circumstellar disks around young stellar objects. Astron. J. 99, 924 (1990).
\item R. C. Bohlin, B. D. Savage, J. F. Drake. A survey of interstellar H I from L-alpha absorption measurements. II. Astrophys. J. 224, 132 (1978).
\item G. Dipierro, P. Pinilla, G. Lodato, and L. Testi. Dust trapping by spiral arms in gravitationally unstable protostellar discs. Mon. Not. R. Astron. Soc., 451, 974 (2015). 
\item D. Foreman-Mackey, D.W. Hogg, D. Lang, J. Goodman. emcee: The MCMC hammer. Publ. Astron. Soc. Pac. 125, 306 (2013).
\item K. A. Rosenfeld, S. M. Andrews, A. M. Hughes, D. J. Wilner, C. Qi. A spatially resolved vertical temperature gradient in the HD 163296 disk. Astrophys. J. 774, 16 (2013).
\item J. L. Rivera, L. Loinard, S. A. Dzib, G. N. Ortiz-Le\'on, L. F. Rodr\'iguez, R. M. Torres. Internal and relative motions of the Taurus and Ophiuchus star-forming regions. Astrophys. J. 807, 119 (2015).
\item J. Menu et al. On the structure of the transition disk around TW Hydrae. Astron. Astrophys. 564, A93 (2014).
\item M. Tazzari et al. Multiwavelength analysis for interferometric (sub-)mm observations of protoplanetary disks. Radial constraints on the dust properties and the disk structure. Astron. Astrophys. 588, A53 (2016).
\item \url{https://science.nrao.edu/facilities/vla/data-processing/pipeline/}
\end{enumerate}

\end{document}